%% file: paper.tex
\documentclass[runningheads]{llncs}
\usepackage[T1]{fontenc}
\usepackage{hyperref}
\usepackage{amsmath}
\usepackage{booktabs}
\usepackage{multirow}
\usepackage{caption}
\usepackage{graphicx}
\usepackage{amsfonts} 
\usepackage{color}
\usepackage{cleveref} 
\usepackage{float}

\usepackage{tikz}
\usepackage{colortbl}      
\usepackage{xcolor}
\usepackage{bbding}
\usepackage{graphicx}
\usepackage{subcaption}  

\definecolor{NaiveColor}{RGB}{0, 114, 178}       
\definecolor{ADcolor}{RGB}{230, 159, 0}       
\definecolor{SDcolor}{RGB}{0, 158, 115}       
\definecolor{CDcolor}{RGB}{204, 121, 167}    
\definecolor{IoUcolor}{RGB}{240, 228, 66}
\definecolor{WholeImageColor}{RGB}{213, 94, 0}     
\definecolor{TheoreticalColor}{RGB}{128, 128, 128}
\definecolor{CLIPcolor}{RGB}{86, 180, 233}  
\definecolor{lightgreen}{RGB}{200, 255, 200}

\newcommand{\MethodBox}[2]{%
  \raisebox{0.5pt}{\textcolor{#1}{\rule{1.2ex}{1.2ex}}}%
  \hspace{0.4ex}\textbf{\textcolor{black}{#2}}%
}

\newcommand{\GNaive}{\MethodBox{NaiveColor}{\textbf{Static}}}
\newcommand{\GAD}{\MethodBox{ADcolor}{\textbf{AD}}}
\newcommand{\GSD}{\MethodBox{SDcolor}{\textbf{SD}}}
\newcommand{\GCD}{\MethodBox{CDcolor}{\textbf{CD}}}
\newcommand{\GIOU}{\MethodBox{IoUcolor}{\textbf{IoU}}}
\newcommand{\GWhole}{\MethodBox{WholeImageColor}{\textbf{Whole-Image}}}
\newcommand{\GTheoretical}{\MethodBox{TheoreticalColor}{\textbf{Theoretical}}}
\newcommand{\GCLIP}{\MethodBox{CLIPcolor}{\textbf{Semantic}}}
\begin{document}
\title{Dynamic Sub-region Search in Homogeneous Collections Using CLIP}
%
%
%
\author{Bastian Jäckl\inst{1}\orcidID{0009-0004-3341-1524} \and
Vojt\v{e}ch Kloda\inst{2}\orcidID{0009-0003-2733-7690} \and Daniel A. Keim\inst{1}\orcidID{0000-0001-7966-9740} \and
Jakub Loko\v{c}\inst{2}\orcidID{0000-0002-3558-4144}}

\institute{University of Konstanz, Konstanz, Germany \and
Charles University, Prague, Czechia}
\authorrunning{Jäckl et al.}
\maketitle              
\begin{abstract}
Querying with text-image-based search engines in highly homogeneous domain-specific image collections is challenging for users, as they often struggle to provide descriptive text queries. For example, in an underwater domain, users can usually characterize entities only with abstract labels, such as corals and fish, which leads to low recall rates. Our work investigates whether recall can be improved by supplementing text queries with position information. Specifically, we explore dynamic image partitioning approaches that divide candidates into semantically meaningful regions of interest. Instead of querying entire images, users can specify regions they recognize. This enables the use of position constraints while preserving the semantic capabilities of multimodal models. We introduce and evaluate strategies for integrating position constraints into semantic search models and compare them against static partitioning approaches. Our evaluation highlights both the potential and the limitations of sub-region-based search methods using dynamic partitioning. Dynamic search models achieve up to double the retrieval performance compared to static partitioning approaches but are highly sensitive to perturbations in the specified query positions.

\end{abstract}

\section{Introduction}
\input{Chapter/1-Introducton}

\section{Related Work} 
\input{Chapter/2-Related-Work}

\section{Sub-region Search Models} \label{sec:searchModels}
\input{Chapter/3-Methods}

\section{Experiments}
\input{Chapter/4-Results}
\section{Conclusions}
\input{Chapter/5-Conclusion}

%
%
\bibliographystyle{splncs04}
\bibliography{paper}

\clearpage
\appendix

\section{Additional experimental results}
In the following tables and figures, we present additional experimental details, which are mentioned in the main text. Thereby, we confirm the key findings stated in the paper.
\subsection{Retrieval effectiveness for using only rectangle distances}
We show in \autoref{tab:only_rects} the retrieval effectiveness using only rectangle distances. Even without using semantic distances, the dynamic grid distances improve effectiveness over semantic distances of the whole-grid and static-grid partitioning.

\begin{table}[H]
    \centering
    \caption{Retrieval performance for Skippable and Non-Skippable annotations using \textbf{only rectangle distances}. The recall is presented as percentage (\%). We highlight best-performing entries.}
    \begin{tabular}{l ||c|c|c|c|c || c|c|c|c|c}
        \toprule
        \multirow{2}{*}{Model} 
        & \multicolumn{5}{c}{Skippable} 
        & \multicolumn{5}{c}{Non-Skippable} \\
        \cmidrule(lr){2-6} \cmidrule(lr){7-11}
        & R@1 & R@10 & R@100 & R@1000 & MNR$\downarrow$ & R@1 & R@10 & R@100 & R@1000 & MNR$\downarrow$  \\
        \midrule
        \midrule
        \multicolumn{11}{l}{\textbf{SAM}} \\
        \GAD         & 0 & 0 & 1 & 6 & 25034 & 0 & 0 & 0 & 6 & 29740 \\
        \GSD          & 0 & 1 & 7 & 20 & 24329 & 0 & 1 & 5 & 14 & 28264 \\
        \GCD          & 0 & 2 & 9 & 28 & 24359 & 0 & 2 & 11 & 24 & 28818 \\
        \GIOU        & 5 & 13 & 23 & 34 & 25861 & 4 & 7 & 17 & 25 & 29031 \\
        \midrule
        \multicolumn{11}{l}{\textbf{Watermask}} \\
        \GAD          & 0 & 0 & 0 & 4 & 22403 & 0 & 0 & 0 & 7 & 27636 \\
        \GSD          & 0 & 1 & 3 & 16 & 16165 & 0 & 0 & 2 & 13 & 24709 \\
        \GCD          & 0 & 1 & 8 & 29 & 12657 & 0 & 1 & 7 & 19 & 19901 \\
        \GIOU         & 2 & 9 & 24 & 46 & 10928 & 1 & 4 & 17 & 30 & 19194 \\
        \midrule
        \multicolumn{11}{l}{\textbf{Grounded}} \\
        \GAD          & 0 & 0 & 0 & 11 & 14387 & 0 & 0 & 1 & 13 & 22143 \\
        \GSD          & 0 & 3 & 11 & 37 & 11692 & 0 & 1 & 8 & 26 & 19879 \\
        \GCD          & 0 & 3 & 21 & 49 & 11691 & 0 & 5 & 18 & 39 & 17993 \\
        \GIOU         & \cellcolor{lightgreen}8 & \cellcolor{lightgreen}25 & \cellcolor{lightgreen}46 & \cellcolor{lightgreen}64 & \cellcolor{lightgreen}10622 & \cellcolor{lightgreen}4 & \cellcolor{lightgreen}18 & \cellcolor{lightgreen}32 & \cellcolor{lightgreen}46 & \cellcolor{lightgreen}16212 \\
        \midrule
    \end{tabular} \label{tab:only_rects}
\end{table}

\subsection{Retrieval effectiveness for using multiplication fusion.}
We show in \autoref{tab:multiplication} retrieval effectiveness for using multiplication fusion of the semantic and rectangle distances. Generally, multiplication fusion works worse than additive fusion.
\begin{table}[h]
    \centering
    \caption{Retrieval performance for Skippable and Non-Skippable annotations using \textbf{multiplication fusion}. The recall is presented as percentage (\%). We highlight best-performing entries.}
    \begin{tabular}{l ||c|c|c|c|c || c|c|c|c|c}
        \toprule
        \multirow{2}{*}{Model} 
        & \multicolumn{5}{c}{Skippable} 
        & \multicolumn{5}{c}{Non-Skippable} \\
        \cmidrule(lr){2-6} \cmidrule(lr){7-11}
        & R@1 & R@10 & R@100 & R@1000 & MNR$\downarrow$ & R@1 & R@10 & R@100 & R@1000 & MNR$\downarrow$  \\
        \midrule
        \midrule
        \multicolumn{11}{l}{\textbf{SAM}} \\
        \GAD          & 0 & 0 & 1 & 6 & 23818 & 0 & 0 & 0 & 6 & 28575 \\
        \GSD          & 0 & 1 & 7 & 21 & 19052 & 0 & 1 & 5 & 15 & 23083 \\
        \GCD          & 0 & 3 & 11 & 30 & 16972 & 1 & 2 & 10 & 23 & 21067 \\
        \GIOU        & 6 & 13 & 24 & 35 & 10253 & 4 & 8 & 17 & 24 & 14361 \\
        \midrule
        \multicolumn{11}{l}{\textbf{Watermask}} \\
        \GAD          & 0 & 0 & 1 & 4 & 22010 & 0 & 0 & 0 & 7 & 27801 \\
        \GSD          & 0 & 1 & 4 & 17 & 15698 & 0 & 0 & 3 & 14 & 24774 \\
        \GCD          & 0 & 2 & 11 & 31 & 12402 & 0 & 1 & 7 & 20 & 20042 \\
        \GIOU         & 3 & 10 & 26 & 48 & 9708 & 1 & 7 & 17 & 30 & 16250 \\
        \midrule
        \multicolumn{11}{l}{\textbf{Grounded}} \\
        \GAD          & 0 & 0 & 0 & 11 & 13421 & 0 & 0 & 1 & 14 & 20628 \\
        \GSD          & 0 & 3 & 12 & 39 & 10728 & 0 & 2 & 9 & 27 & 18240 \\
        \GCD          & 0 & 3 & 21 & 53 & 10359 & 0 & 4 & 20 & 41 & 15831 \\
        \GIOU         & \cellcolor{lightgreen}11 & \cellcolor{lightgreen}27 & \cellcolor{lightgreen}47 & \cellcolor{lightgreen}68 & \cellcolor{lightgreen}8299 & \cellcolor{lightgreen}4 & \cellcolor{lightgreen}20 & \cellcolor{lightgreen}32 & \cellcolor{lightgreen}47 & \cellcolor{lightgreen}11161 \\
        \midrule
    \end{tabular} \label{tab:multiplication}
\end{table}

\subsection{Retrieval effectiveness for using short text queries}
We show the retrieval effectiveness for addititive fusion and \emph{short} text queries in \autoref{tab:overview_short}. Overall, the performance drops are moderately substantantial compared to long text queries.
 \begin{table}[H]
    \centering
    \caption{Retrieval performance for Skippable and Non-Skippable annotations for \textbf{short text queries}. The recall is presented as percentage (\%). We highlight best-performing entries.}
    \resizebox{\textwidth}{!}{
    \begin{tabular}{l ||c|c|c|c|c || c|c|c|c|c}
        \toprule
        \multirow{2}{*}{Model} 
        & \multicolumn{5}{c}{Skippable} 
        & \multicolumn{5}{c}{Non-Skippable} \\
        \cmidrule(lr){2-6} \cmidrule(lr){7-11}
        & R@1$\uparrow$& R@10$\uparrow$& R@100$\uparrow$& R@1000$\uparrow$& MNR$\downarrow$ & R@1$\uparrow$& R@10$\uparrow$& R@100$\uparrow$& R@1000$\uparrow$& MNR$\downarrow$  \\
        \midrule
        \GWhole       & 0 & 3 & 15 & 40 & 8861 & 0 & 1 & 10 & 28 & 13242 \\
        \midrule
        \GNaive       & 0 & 4 & 19 & 51 & 7703 & 1 & 4 & 12 & 31 & 9555 \\
        \midrule
        \multicolumn{11}{l}{\textbf{SAM}} \\
        \GCLIP          & 0 & 4 & 15 & 45 & 7383 & 0 & 1 & 9 & 29 & 13053 \\
        \GAD         & 1 & 5 & 18 & 36 & 12219 & 0 & 1 & 5 & 22 & 19029 \\
        \GSD          & 2 & 8 & 25 & 43 & 10120 & 0 & 2 & 9 & 27 & 16698 \\
        \GCD          & 2 & 12 & 30 & 48 & 8721 & 1 & 3 & 13 & 31 & 13457 \\
        \GIOU         & 13 & 21 & 34 & 50 & 6830 & 3 & 12 & 18 & 35 & 11382 \\
        \midrule
        \multicolumn{11}{l}{\textbf{Watermask}} \\
        \GCLIP          & 0 & 3 & 18 & 50 & 7849 & 0 & 1 & 11 & 31 & 11722 \\
        \GAD          & 1 & 5 & 23 & 55 & 7181 & 0 & 3 & 10 & 32 & 14852 \\
        \GSD          & 1 & 8 & 30 & 61 & 6274 & 0 & 3 & 15 & 35 & 13995 \\
        \GCD          & 4 & 14 & 38 & 65 & 5640 & 2 & 8 & 20 & 41 & 10816 \\
        \GIOU         & 13 & 32 & 52 & 69 & \cellcolor{lightgreen}5520 & 5 & 17 & 32 & 47 & 11147 \\
        \midrule
        \multicolumn{11}{l}{\textbf{Grounded}} \\
        \GCLIP          & 0 & 3 & 17 & 49 & 8332 & 0 & 1 & 10 & 32 & 11421 \\
        \GAD          & 1 & 6 & 23 & 60 & 7276 & 0 & 2 & 14 & 34 & 14019 \\
        \GSD          & 2 & 10 & 34 & 64 & 6498 & 0 & 5 & 16 & 36 & 11683 \\
        \GCD         & 3 & 17 & 44 & 66 & 6408 & 3 & 9 & 22 & 43 & 9714 \\
        \GIOU         & \cellcolor{lightgreen}23 & \cellcolor{lightgreen}42 & \cellcolor{lightgreen}62 & \cellcolor{lightgreen}74 & 6330 & \cellcolor{lightgreen}11 & \cellcolor{lightgreen}25 & \cellcolor{lightgreen}37 & \cellcolor{lightgreen}51 & \cellcolor{lightgreen}8589 \\
        \midrule
        \GTheoretical & 5 & 10 & 28 & 56 & 8120 & 2 & 5 & 18 & 41 & 9118 \\
        \bottomrule
    \end{tabular}} \label{tab:overview_short}
\end{table} 

\subsection{Robustness to Perturbation}
\begin{figure}[h]
    \centering
    \includegraphics[width=1\linewidth]{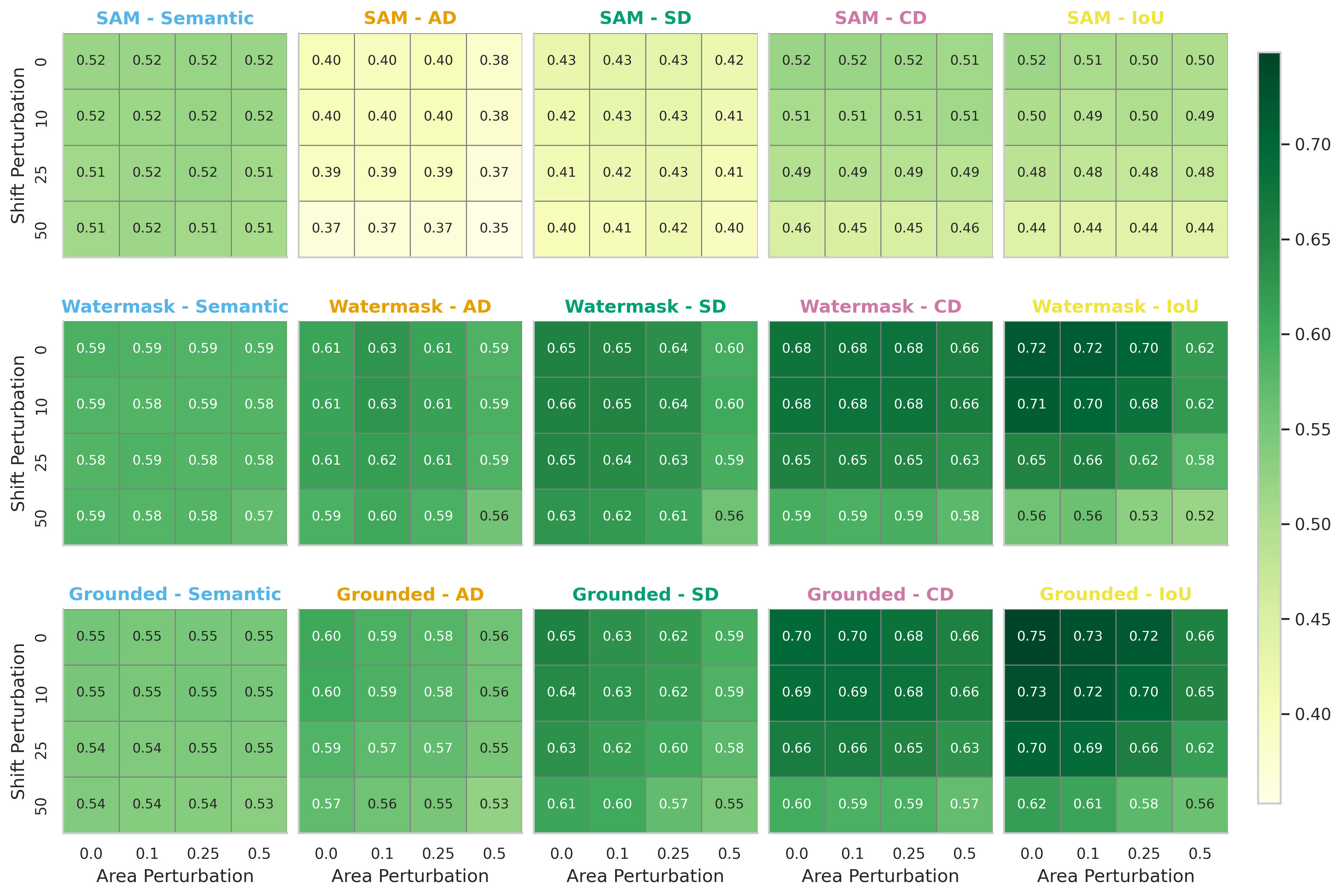}
    \caption{Heatmap showing R@1000 for (model, rectangle-distance) pairs for skippable annotations.}
    \label{fig:perturbations_1000}
\end{figure}

We show the robustness of rectangle distances for recall@1000 in \autoref{fig:perturbations_1000}. For r@1000, all grids show significantly better robustness compared to r@100. 

\end{document}

%% file: Chapter/1-Introducton.tex
With the emergence of powerful multimodal search models \cite{clip}, querying by text in vast image/video collections has become incredibly effective. The reason is that users can usually specify highly descriptive text queries, whose semantics are automatically embedded and matched against multimedia objects in the database. However, in highly homogeneous domains, such as underwater imagery or medical datasets, this approach breaks down \cite{vbs2024_eval_performance}. Users in these settings often struggle to formulate sufficiently descriptive queries, relying instead on abstract terms, leading to large candidate sets and low retrieval precision \cite{staticApproach}. Consequently, specifying object regions might be the only clue to improving the discriminative capabilities between images containing similar abstract entities. These observations motivate the need for retrieval systems that incorporate both semantic and spatial signals.

\begin{figure}
\centering
\begin{tikzpicture}[scale=1]
    \node at (-0.5,3.25) {$k_1$};
    \node at (-0.5,1.05) {$k_2$};
    \draw[thick] (0,0) rectangle (3,2);
    \draw[dashed,lightgray] (0.75,0.5) rectangle (2.25,1.5);
    \draw[dashed] (1.5,0) -- (1.5,2);
    \draw[dashed] (0,1) -- (3,1);
    
    \draw[thick] (0,2.2) rectangle (3,4.2);
    \draw[dashed] (1.5,2.2) -- (1.5,4.2);
    \draw[dashed] (0,3.2) -- (3,3.2);
    \draw[dashed,lightgray] (0.75,2.7) rectangle (2.25,3.7);
    \node at (1.3,-0.5) {a) Static Partitioning};

    \draw[thick] (4,0) rectangle (7,2);
    \draw[olive,dashed] (4.2,1.1) rectangle (6.2,1.7);
    \node[olive] at (4.7,3.5) {query};
    \draw[blue,dashed] (4.3,0.2) rectangle (5.3,0.7);
    \node[blue] at (4.8,2.6) {query};
    \draw[teal,dashed] (5.5,0.3) rectangle (6.7,1.8);
    \node[teal] at (6.1,2.7) {query};
    \draw[thick] (4,2.2) rectangle (7,4.2);
    \draw[olive,dashed] (4.2, 3.3) rectangle (6.2,3.9);
    \draw[blue,dashed] (4.3,2.4) rectangle (5.3,2.9);
   \draw[teal,dashed] (5.5,2.5) rectangle (6.7,4);
    \node at (5.4,-0.5) {b) Theoretical Optimal};

    \draw[thick] (8,0) rectangle (11,2);
    \draw[dashed] (8.2,0.1) rectangle (9,1);
    \draw[dashed] (8.6,1.2) rectangle (9.2,1.7);
    \draw[dashed] (9.4,1.1) rectangle (10.2,1.3);
    \draw[dashed] (9.3, 0.2) rectangle (10.1,0.6);
    \draw[dashed] (9.8, 0.3) rectangle (10.3,0.8);
    
    \draw[thick] (8,2.2) rectangle (11,4.2);
    \draw[dashed] (8.5,2.4) rectangle (9,2.8);
    \draw[dashed] (9.1,2.3) rectangle (9.4,2.9);
    \draw[dashed] (9.5,2.5) rectangle (9.9,3);
    \draw[dashed] (10,3.5) rectangle (10.4,3.9);
    \draw[dashed] (10.5,3.3) rectangle (10.8,3.6);
    \node at (9.3,-0.5) {c) Dynamic Partitioning};

\end{tikzpicture}
    \caption{Examples of image partitioning options for two database images: a) static partitioning to five regions with four corner and one center region, b) theoretical partitioning illustrated for three (out of many possible) query regions specified by users, and c) dynamic partitioning to five regions based on automatic region detection.}
    \label{fig:example}
\end{figure}

To address this, we investigate whether region-constrained retrieval, where users specify both a textual description and a spatial region, can significantly improve search effectiveness. Existing approaches that use static partitioning, such as dividing images into fixed grids, enable basic sub-region search but ignore the semantic structure of images, often splitting objects or merging unrelated content \cite{LokocMVS21,staticApproach}. In this work, we explore dynamic sub-region search relying on object detection and segmentation models to automatically partition images into coherent, meaningful regions (see Figure~\ref{fig:example} for a comparison). The search system encodes extracted regions using CLIP and integrates position-aware constraints, such as bounding box alignment, into the similarity ranking process. We evaluate various strategies for fusing semantic and spatial signals. Crucially, we also compare our dynamic approach to both static partitioning and a theoretical optimal cropping, relying directly on ground-truth annotations. Our experiments thereby focus on upper-bound performance estimates. For this purpose, we use a set of ground-truth annotations, collected for keyframes of the Marine Video Kit (MVK) \cite{MVK}, containing highly homogenoeus underwater videos. The annotations consist of image sub-region boundaries and their text descriptions. We analyze the effect of query region perturbations with artificial white noise (simulating region specification inaccuracy) as well. Summarizing, our key findings are:

\begin{itemize}
    \item Dynamic sub-region search models that combine semantic and spatial information significantly outperform static grid-based approaches.
    \item In contrast, models that use dynamic segmentation matching but ignore region constraints in the distance function do not improve retrieval performance.
    \item Dynamic methods are sensitive to region specification errors, they still outperform traditional baselines under realistic perturbations.
\end{itemize}


%% file: Chapter/2-Related-Work.tex
Our work contributes to the field of sub-region image search in the highly homogeneous domain of underwater images. We review advances in both fields.

\subsection{Sub-Image Search}
Applying sub-region search for content-based image queries is a long-established technique \cite{visualSeek}. Early approaches primarily leverage low-level features such as color \cite{visualSeek,signatureVideoBrowser} where grid-partitionings directly derived from these low-level features. Later, automatic object detectors were applied to build systems that allow for dynamic grids \cite{objectGraph,regionBasedRetrieval,spatialSemanticSearch} respecting semantic coherence. These early systems relied on building queries with a predefined set of labels, not allowing for complex queries in natural language. An exception is Xu et al. \cite{conceptMap}, proposing an image search system that enables users to specify both semantic concepts and their spatial layout by formulating a word-based concept map on a canvas, which is then translated into a visual instance map by querying another web-based image search engine to find representatives of the individual concepts. 

Powerful Vision Foundation Models (VFM), predominantly CLIP \cite{clip}, changed the landscape of image-retrieval systems. Nowadays, most of them rely on semantic text queries. However, naively appending position information to textual queries in VFMs leads to unsatisfactory performance \cite{badLocalization}, especially in homogeneous domains \cite{staticApproach}. Consequently, researchers explored more elaborate pipelines based on VFMs: Shlapentokh-Rothman et al. \cite{regionBasedRepresentations} revisit region-based image representations by combining SAM-generated masks \cite{kirillov2023segany} with DINOv2 features \cite{dinov2}, showing that simple pooled region features can achieve competitive performance in tasks like semantic segmentation and object-based retrieval, while enabling efficient and customizable image search. The closest work to our approach is Search Anything \cite{searchAnything}, a zero-shot, prompt-based image region retrieval method that combines FastSAM segmentation \cite{fastSam} and CLIP semantics to learn binary hash codes for fine-grained, region-level similarity search without relying on labeled data. However, Search Anything requires image queries, while our work utilizes the cross-modal capability of CLIP to query by text. We furthermore compare our dynamic approach to static baselines, investigated by Jäckl et al. for underwater images \cite{staticApproach}.

\subsection{Region Segmentation for Underwater Images}
Dynamic sub-region search models require partitionings of images that accurately match (previously unknown) queries. Our basic assumption is that users query for specific \emph{objects} or semantically similar \emph{regions}. We thus rely on an accurate detection of these regions. Thereby, we apply direct object detectors or segmentation models whose masks can be naively transformed into bounding boxes. However, training deep learning networks is challenging for the underwater domain, as training data is scarce \cite{suim,fishDataset}. Most of the existing object detectors are thus limited to specific classes of underwater life \cite{fishDataset,reviewUnderwater}. Only recently, Lian et al. \cite{watermask} published a larger-scale dataset with 4628 images with pixel-level annotated masks of seven categories (fish, reefs, aquatic plants, wrecks/ruins, divers, robots, and sea floor). This allowed them to train the instance segmentation model WaterMask, which advances the state-of-the-art. We will use WaterMask as a representative for specialized deep-learning models.

As an alternative, we investigate the applicability of VFMs trained on large-scale datasets for object detection or segmentation. Specifically, we deploy SAM2 \cite{ravi2024sam2segmentimages}, which creates segmentation of whole images. It was trained on the SA-V dataset containing 35.5M masks across 50.9K videos. Essential for our work is that SAM2 is an open-set model that can even segment object categories that are not included in the training set - other object detectors, such as YOLO \cite{yolov11}, are not practical for our work, as they are trained to detect specific object categories that do not match the underwater scenery. The third model we apply is Grounded-SAM2 \cite{sam2github}, which combines the segmentation capabilities of SAM2 with the open-set object detector Grounding-Dino \cite{liu2023grounding,ren2024grounded}. Grounding-Dino is a zero-shot object detector that uses natural language prompts to localize and identify arbitrary objects in images without requiring task-specific training. Consequently, Grounded-SAM2 allows us to formulate which objects/regions should be detected by the model.

%% file: Chapter/3-Methods.tex
We use this chapter to formally introduce sub-region search models. We start by introducing the commonly used, region-unaware \GWhole\ baseline. This system utilizes CLIP to embed keyframes \( \{k_i\}_{i=1}^n \) into features \( F_K \in \mathbb{R}^{n \times d} \), where each row \( f_{k_i} \in \mathbb{R}^d \) represents the embedding of keyframe \( k_i \). At query time, a text query \( t \) is encoded into a feature vector \( f_t \in \mathbb{R}^d \) using CLIP’s text encoder. Assuming all vectors are \( \ell_2 \)-normalized, similarity scores \( \mathbf{s} \in \mathbb{R}^n \) are computed as:

\[
\mathbf{s} = F_K f_t^\top
\]

Each entry \( s_i = \langle f_{k_i}, f_t \rangle \) corresponds to the cosine similarity between the query and keyframe \( k_i \), allowing keyframes to be ranked by semantic relevance.

\subsection{Static Partitioning}
\GWhole\ relies on highly discriminative text queries, which are not always available in practice. Consequently, researchers combined text queries with localized information for image retrieval, showing tremendous success. Our work is an extension of Jäckl et al. \cite{staticApproach}, which established \GNaive\ grid partitioning for the challenging domain of underwater scenery. We will first introduce this \GNaive\ approach and subsequently extend it to dynamic methods.

Sub-region-based image retrieval requires a textual description and position information. For the purpose of this paper, we assume that we have a query consisting of text description \( t \) and an annotation bounding box \( b \). \GNaive\ image retrieval then proceeds in two stages. The first stage selects candidate subregions from each image. Therefore, we define a static grid over each image and consider as candidates all grid cells \( r \in R \) that intersect with the annotation bounding box \( b \). This yields a set of candidate regions

\[
R_b = \{ r \in R \mid \text{IoU}(r, b) > 0 \}.
\]

The intersection-over-union (IoU) between two axis-aligned rectangular regions can be computed in constant time, allowing for efficient region selection. While this setup considers all overlapping regions \( R_b \), an alternative is to select only the region with the highest IoU per image:

\[
r^* = \arg\max_{r \in R} \text{IoU}(r, b).
\]

In the second stage, each candidate region \( r \in R_b \) is associated with a precomputed CLIP image embedding \( f_r \in \mathbb{R}^d \), and the text query \( t \) is embedded as \( f_t \in \mathbb{R}^d \) using CLIP’s text encoder. Cosine similarity scores are similarly computed as the dot product between $f_t$ and every $f_r$ yielding a ranking over all subregions. Final image-level ranks are derived by selecting, for each image, the subregion with the highest similarity score (i.e., the lowest rank). For our experiments, we use the 5-grid setting introduced by Jäckl et al. \cite{staticApproach} (see \autoref{fig:example}, c). They also experimented with a 9-grid partitioning, which showed numbers similar to those of the 5-grid setting.

\subsection{Dynamic Region-Based Retrieval with Geometric Constraints} \label{sec:pipeline}

\begin{figure}
    \centering
    \includegraphics[width=1\linewidth]{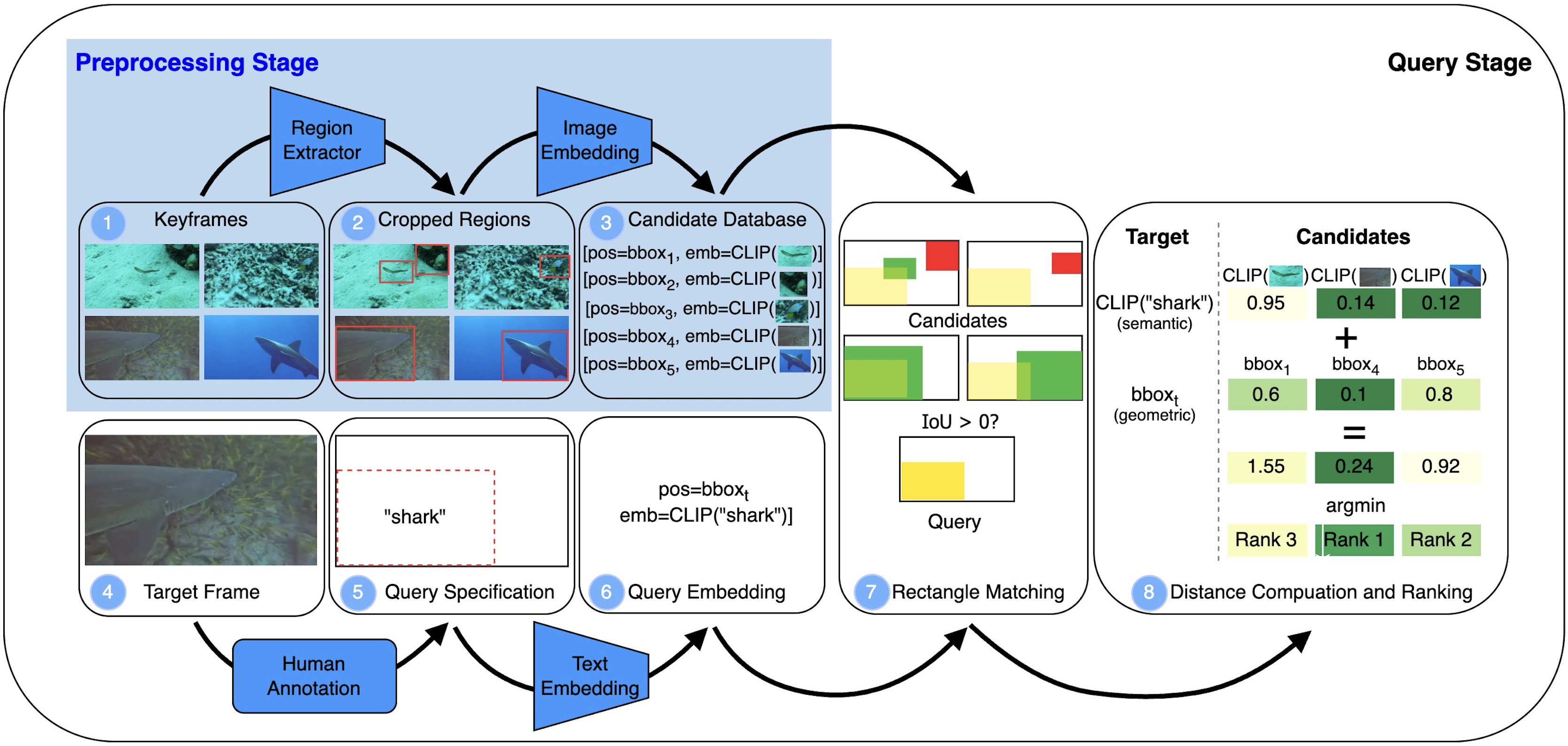}
    \caption{\textbf{Pipeline for dynamic region-based image retrieval}. The individual steps are described in the main text.}
    \label{fig:pipeline}
\end{figure}

The \GNaive\ grid-based approach described earlier does not account for semantic structure within images, often resulting in subregions that divide objects or include excessive background. To address this, we propose dynamic, content-aware partitioning by applying automatic segmentation. An overview of this approach is given in \autoref{fig:pipeline}. Our approach is distributed into two phases (preprocessing and query time).

\subsubsection{Preproccesing Stage}
The preprocessing stage  in \autoref{fig:pipeline}) begins with non-annotated images (or keyframes for video retrieval, \textbf{Step 1}). \textbf{Step 2} automatically extracts regions of interest that should align with human perceptions. Prior work has achieved this either through models trained on large-scale supervised training data on general-domain datasets~\cite{spatialSemanticSearch} or by leveraging low-level image cues~\cite{visualSeek}. In contrast, our approach targets a highly specialized and homogeneous domain—underwater scenes—where annotated data is scarce. We, therefore, rely on open-set foundation models capable of segmenting arbitrary visual scenes without requiring task-specific supervision. This allows us to generate a set of semantically coherent regions \( \{r_j\}_{j=1}^{m_i} \) for each image \( I_i \). Each region \( r_j \subset I_i \) is defined by a bounding box \( \text{rect}_j = (\text{left}_j, \text{top}_j, \text{width}_j, \text{height}_j) \). We use these bounding boxes in \textbf{Step 3}, where we construct CLIP embeddings \( f_{r_j} \in \mathbb{R}^d \) for the regions of interest. They are stored in a database with the respective bounding boxes.

\subsubsection{Query Time Stage}
During query time, the objective is to retrieve the most relevant target from the database based on a user-provided query (\textbf{Step 4}). To this end, the user specifies a query comprising a textual description \( t \) and an associated annotation in the form of a bounding box \( b \) (\textbf{Step 5}). The textual description \( t \) is embedded into a feature vector \( f_t \in \mathbb{R}^d \) using the CLIP text encoder (\textbf{Step 6}).

To identify relevant candidates, the system first retrieves all indexed subregions whose bounding boxes overlap with the query region \( b \), analogous to static approaches. Specifically, the set of candidate regions \( R_b \) is defined as:

\[
R_b = \left\{ r_j \;\middle|\; \text{IoU}(\mathrm{rect}_j, b) > 0 \right\},
\]

where \( r_j \) denotes a segmented region with bounding box \( \mathrm{rect}_j \).

In (\textbf{Step 8}), the remaining candidate regions are ranked based on their similarity to the query. Given the embedded text query \( f_t \), we compute the semantic similarity between \( f_t \) and each region embedding \( f_{r_j} \) using cosine similarity:

\[
s_j = \left\langle f_{r_j}, f_t \right\rangle.
\]

This score \( s_j \) is subsequently fused with a geometric distance measure to rank all candidate regions for final retrieval. Therefore, we define geometric distance functions between two rectangles \( \text{rect}_A \) and \( \text{rect}_B \) as follows:

\begin{itemize}
    \item \textbf{Area Distance (\GAD)}: Absolute difference of areas:
    \[
    D_{\text{AD}} = \left| \text{area}_A - \text{area}_B \right|.
    \]
    \item \textbf{Shape Distance (\GSD)}: Sum of width and height differences:
    \[
    D_{\text{SD}} = \left| \text{width}_A - \text{width}_B \right| + \left| \text{height}_A - \text{height}_B \right|.
    \]
    \item \textbf{Centroid Distance (\GCD)}: Measures Euclidean distance between centers:
    \[
    D_{\text{CD}} = \left\|c_A - c_B\right\|_2, \quad \text{where } c_A = \left(\text{left}_A + \tfrac{1}{2}\text{width}_A, \text{top}_A + \tfrac{1}{2}\text{height}_A\right).
    \]
    \item \textbf{Intersection-over-Union (\GIOU)}: Measures overall overlap:
    \[
    \text{IoU}(\text{rect}_A, \text{rect}_B) = \frac{|\text{rect}_A \cap \text{rect}_B|}{|\text{rect}_A \cup \text{rect}_B|}, \quad \text{with distance } D_{\text{IoU}} = 1 - \text{IoU}.
    \]
\end{itemize}

The distance functions vary in their type of geometric constraints. While \GAD\ only compares the sizes, \GSD\ compares the size and aspect ratio and is thus a stricter constraint. \GCD\ neglects shape differences but considers the position on the grid. \GIOU\ is the strongest constraint, considering the area, shape, and position of rectangles. Let \( D_{\text{rect}} \in \mathbb{R} \) be any of the above geometric distance functions. We combine it with the cosine-based CLIP distance \( D_{\text{CLIP}} = 1 - \langle f_{r_j}, f_t \rangle \) to form a joint distance score. We standardize all distance functions to \( \mathcal{N}(0, 1) \). We consider two fusion strategies:

\begin{itemize}
    \item \textbf{Linear Combination}:
    \[
    D_{\text{combined}} = (1 - \alpha) \cdot D_{\text{CLIP}} + \alpha \cdot D_{\text{rect}}, \quad \alpha \in [0, 1].
    \]
    \item \textbf{Geometric Mean Combination}:
    \[
    D_{\text{combined}} = D_{\text{CLIP}}^{(1 - \alpha)} \cdot D_{\text{rect}}^{\alpha}, \quad \alpha \in [0, 1].
    \]
\end{itemize}

As a starting point, this paper considers only $\alpha = 0.5$ for experiments, rating both distances equally important. We note that with other $\alpha$-values, the dynamic approaches could even improve more.

We additionally examine the baseline approach using only CLIP similarity (i.e., without geometric constraints), denoted as \GCLIP.

\subsection{Theoretical Upper-Bound Baseline}
We furthermore use a upper-bound baseline established by Jäckl et al. \cite{staticApproach}, which defines a variant that uses the annotated bounding boxes \( \{b_i\} \) directly to construct a new version of the dataset. Each frame is cropped to the exact coordinates of its annotated object bounding box \( b_i = (\text{left}_i, \text{top}_i, \text{width}_i, \text{height}_i) \), resulting in a derived dataset \( \mathcal{D}_{\text{crop}} \) for each annotation containing tightly focused object regions. For each cropped region, we compute its CLIP embedding \( f_{b_i} \in \mathbb{R}^d \), and compare it to the query embedding \( f_t \in \mathbb{R}^d \) via cosine distance, similar to the other grid approaches. This setup (\GTheoretical) mimics the scenario where the frame is partitioned with infinitely fine granularity and the optimal region is always selected. While computationally impractical for real-time systems, this oracle baseline serves as an upper bound on retrieval performance given perfect region localization.

%% file: Chapter/4-Results.tex
Our work explores the potential of \emph{dynamic} sub-region search in the highly homogeneous underwater scenery. We chose this specific domain as it has proven challenging in competitive retrieval contests, such as the Video Browser Showdown (VBS) \cite{vbs2024_eval_performance,vbs2023,vbsEval}. Specifically, we use a set of keyframes extracted from the MarineVideoKit (MVK) ~\cite{MVK} to create the candidate database. Those were used by teams participating in the VBS \cite{stroh2025prak}. As queries, we apply a set of human-annotated keyframes, including short and long textual descriptions and bounding boxes for regions of interest, collected by Jäckl et al \cite{staticApproach}. These annotations provide \emph{optimal bounding boxes}, as annotators could add bounding boxes directly on images. The presented results in \autoref{sec:effectiveness} are thus theoretical upper bounds. Furthermore, the annotations are partitioned in subsets of "Skippable" and "Non-Skippable" - in "Skippable" annotations, annotators could choose which image they want to annotate, while they had to annotate all (randomly chosen) images for "Non-Skippable", even if there was no apparent region of interest. As a basic metric, we measure the rank of the human-annotated frame in the keyframe database. We report further ablations with ill-positioned bounding boxes in \autoref{sec:shifts}. While Jäckl et al. showed performance gains for static grid partitioning, we primarily explore approaches that partition the grid dynamically with the help of segmentation and detection models.

\subsection{Retrieval Effectiveness}\label{sec:effectiveness}
We report the recall (R@X) and average rank (MNR) of search models introduced in \autoref{sec:searchModels} in \autoref{tab:overview}. The obtained numbers are generated using the long textual annotation with linear fusion. We report the results for geometric mean fusion in \autoref{tab:multiplication}. However, we observe consistently better results with linear fusion.

\begin{table}[ht]
    \centering
    \caption{Retrieval performance for Skippable and Non-Skippable annotations. The recall is presented as percentage (\%). We highlight best-performing entries.}
    \resizebox{\textwidth}{!}{
    \begin{tabular}{l ||c|c|c|c|c || c|c|c|c|c}
        \toprule
        \multirow{2}{*}{Model} 
        & \multicolumn{5}{c}{Skippable} 
        & \multicolumn{5}{c}{Non-Skippable} \\
        \cmidrule(lr){2-6} \cmidrule(lr){7-11}
        & R@1$\uparrow$& R@10$\uparrow$& R@100$\uparrow$& R@1000$\uparrow$& MNR$\downarrow$ & R@1$\uparrow$& R@10$\uparrow$& R@100$\uparrow$& R@1000$\uparrow$& MNR$\downarrow$  \\
        \midrule
        \GWhole       & 0 & 6 & 21 & 47 & 9320 & 0 & 2 & 11 & 34 & 11064 \\
        \midrule
        \GNaive       & 1 & 9 & 29 & 56 & 7715 & 0 & 2 & 18 & 43 & 7419 \\
        \midrule
        \multicolumn{11}{l}{\textbf{SAM}} \\
        \GCLIP          & 1 & 7 & 25 & 52 & 7880 & 0 & 3 & 13 & 33 & 11773 \\
        \GAD         & 2 & 7 & 22 & 40 & 11924 & 0 & 4 & 11 & 25 & 17810 \\
        \GSD          & 3 & 11 & 26 & 43 & 10243 & 0 & 4 & 14 & 29 & 15619 \\
        \GCD          & 5 & 15 & 34 & 52 & 8429 & 2 & 5 & 18 & 35 & 12834 \\
        \GIOU         & 14 & 22 & 34 & 52 & 7300 & 5 & 15 & 21 & 40 & 10781 \\
        \midrule
        \multicolumn{11}{l}{\textbf{Watermask}} \\
        \GCLIP          & 1 & 11 & 30 & 59 & 6585 & 0 & 4 & 13 & 40 & 9229 \\
        \GAD          & 3 & 12 & 34 & 61 & 6544 & 1 & 4 & 17 & 42 & 13897 \\
        \GSD          & 4 & 16 & 38 & 65 & 5777 & 1 & 6 & 21 & 44 & 12603 \\
        \GCD          & 7 & 22 & 45 & 68 & 5281 & 2 & 11 & 26 & 50 & 9819 \\
        \GIOU         & 16 & 36 & 55 & 72 & \cellcolor{lightgreen}5022 & 8 & 21 & 35 & 51 & 9728 \\
        \midrule
        \multicolumn{11}{l}{\textbf{Grounded}} \\
        \GCLIP          & 2 & 9 & 29 & 55 & 7590 & 0 & 3 & 11 & 39 & 9296 \\
        \GAD          & 2 & 11 & 34 & 60 & 6867 & 0 & 6 & 17 & 39 & 13265 \\
        \GSD          & 6 & 17 & 40 & 65 & 6375 & 1 & 7 & 21 & 42 & 10654 \\
        \GCD         & 7 & 25 & 47 & 70 & 5808 & 2 & 11 & 28 & 49 & 9114 \\
        \GIOU         & \cellcolor{lightgreen}25 & \cellcolor{lightgreen}46 & \cellcolor{lightgreen}62 & \cellcolor{lightgreen}75 & 5856 & \cellcolor{lightgreen}13 & \cellcolor{lightgreen}29 & \cellcolor{lightgreen}40 & \cellcolor{lightgreen}55 & 7765 \\
        \midrule
        \GTheoretical & 8 & 19 & 41 & 67 & 5684 & 5 & 11 & 30 & 57 & \cellcolor{lightgreen}5447 \\
        \bottomrule
    \end{tabular}} \label{tab:overview}
\end{table}

\subsubsection{Dynamic Baseline}
\GCLIP\ relies on matching dynamically extracted bounding boxes but only employs semantic distance for ranking. For skippable annotations, \GCLIP\ yields only marginal improvements over \GNaive, and is even worse for non-skippable annotations. Surprisingly, these outcomes persist despite a high measured overlap between the annotated and predicted rectangles. For instance, while the average IoU between the best-matching rectangle in the target frame and the corresponding annotation for Grounded-SAM (under skippable annotations) is 0.61 (using 6.34 rectangles per frame on average), the respective IoU for \GNaive\ is only 0.31 (using five rectangles) —yet both approaches achieve similar retrieval performance. Likewise, the IoU for Watermask is 0.61 (using 13.74 rectangles) and 0.40 for SAM (using 7.68 rectangles). These results suggest that the lack of performance gains cannot be attributed to poor region detection. 

We analyze two additional factors that explain why \GCLIP\ fails to outperform \GNaive. First, while IoU captures alignment quality, it does not indicate whether an annotation is coherently represented by a single region. This is especially relevant for small annotations, which bear a low IoU when using static grid cells. Thus, we further compute the maximum coverage of each annotation by any single candidate region and report the average across all annotations. The results show only marginal differences: 63\% for \GCLIP\ using Grounded, and 57\% for \GNaive. This suggests that although \GNaive\ includes more unannotated context, it still captures the annotated content with similar coherence.
Second, we examine whether region detectors help exclude irrelevant areas. To this end, we measure the proportion of the frame that is considered during ranking, defined as the union of regions satisfying the matching criterion $\text{IoU} > 0$. Interestingly, dynamic approaches such as \GCLIP\ cover even more of the frame (70\%) than the static grid used in \GNaive\ (51\%). This indicates that region detectors can not effectively filter out seemingly uninteresting regions. Even if we take best IoU per frame, we still have no improvement over static approach. At the same time, the larger bounding boxes in static grids do not significantly degrade embedding quality. We provide evidence for the latter by combining the similarity scores extracted by \GCLIP\ (Grounded) and \GNaive. Due to a violation of the normality assumption (Shapiro–Wilk test, $p < 0.05$), we use a Wilcoxon signed rank to test for statistical differences. The test did not reveal a statistically significant difference ($p = 0.169$). A Pearson-Correlation Coefficient of 0.63 between the ranks observed by \GCLIP\ and \GNaive\ also confirms the strong relation.

Consequently, we notice that a tight spatial alignment is insufficient to rank frames accurately. The \GTheoretical\ baseline further supports this interpretation. Even with an infinitely fine-grained grid, only 57\% of target items (in the non-skippable case) appear among the top 1000 results. While such a fine-grained grid improves retrieval coverage, it also demonstrates the limitations of using CLIP similarity alone as a ranking signal. High spatial resolution does not guarantee successful retrieval if the semantic match to the text query is weak or ambiguous. In the next subsection, we thus explore whether the strong spatial alignments can be used for more effective retrieval.

\subsubsection{Combining Semantic and Geometric Ranking} Since ranking functions based solely on text queries result in unsatisfactory performance, we explicitly incorporate geometric constraints from the annotation rectangle into the \emph{ranking} function. The experiments presented were conducted using perfect query bounding boxes, representing the upper bounds of dynamic grid partitionings. As shown in \autoref{tab:overview}, augmenting the baseline \GCLIP\ with area distance (\GAD) yields only slight improvements in effectiveness. More substantial gains are seen with \GSD\ and \GCD\, while \GIOU\ delivers the most significant leap, even surpassing the \GTheoretical\ baseline.

At first glance, this result may seem counterintuitive. After all, \GTheoretical\  uses a dataset specifically cropped for each annotation, which should theoretically provide a strong advantage. However, it does not respect geometrical constraints, allowing text queries to match with everything inside the box. This is especially problematic for big annotation boxes. In contrast, the \GIOU\ method not only verifies whether a relevant object appears in the region but also \emph{precisely} evaluates the alignment of shape and position. Remarkably, even when rankings are based solely on geometric distances, \GIOU\ significantly outperforms \GTheoretical, as shown in \autoref{tab:only_rects_short}. Similarly, we observe that using only rectangles as distance leads to worse performance than the semantic and geometry distance fusion.
This effect is also visible in the rank distributions for short text queries (Appendix \autoref{tab:overview_short}): While \GTheoretical\ performance drops sharply, the rectangle-based distances maintain consistent effectiveness. Those results utilize perfect annotations- we assess how those work in practice in the next section.

\begin{table}[ht]
    \centering
    \caption{Retrieval performance for Skippable and Non-Skippable annotations using only rectangle distances. We highlight best-performing entries.}
    \resizebox{\textwidth}{!}{
    \begin{tabular}{l ||c|c|c|c|c || c|c|c|c|c}
        \toprule
        \multirow{2}{*}{Model} 
        & \multicolumn{5}{c}{Skippable} 
        & \multicolumn{5}{c}{Non-Skippable} \\
        \cmidrule(lr){2-6} \cmidrule(lr){7-11}
        & R@1$\uparrow$& R@10$\uparrow$& R@100$\uparrow$& R@1000$\uparrow$& MNR$\downarrow$ & R@1$\uparrow$& R@10$\uparrow$& R@100$\uparrow$& R@1000$\uparrow$& MNR$\downarrow$  \\
        \midrule
        \midrule
        \multicolumn{11}{l}{\textbf{Grounded}} \\
        \GAD          & 0 & 0 & 0 & 11 & 14387 & 0 & 0 & 1 & 13 & 22143 \\
        \GSD          & 0 & 3 & 11 & 37 & 11692 & 0 & 1 & 8 & 26 & 19879 \\
        \GCD          & 0 & 3 & 21 & 49 & 11691 & 0 & 5 & 18 & 39 & 17993 \\
        \GIOU         & \cellcolor{lightgreen}8 & \cellcolor{lightgreen}25 & \cellcolor{lightgreen}46 & \cellcolor{lightgreen}64 & \cellcolor{lightgreen}10622 & \cellcolor{lightgreen}4 & \cellcolor{lightgreen}18 & \cellcolor{lightgreen}32 & \cellcolor{lightgreen}46 & \cellcolor{lightgreen}16212 \\
        \bottomrule
    \end{tabular}} \label{tab:only_rects_short}
\end{table}

\subsection{Robustness to bounding box shifts}\label{sec:shifts}
Previously presented experiments are upper bounds for dynamic approaches with optimally aligned annotation boxes. We now assess their robustness to perturbations as they may occur in realistic scenarios. Therefore, we apply two kinds of perturbations: 1) we shift the box's position, and 2) we modify its size. For the shifts, we draw $ d_{shiftX} $ and $d_{shiftY}$ independently for each annotation from a normal distribution: $ d_{shiftX},d_{shiftY} \sim \mathcal{N}(0, \sigma_s)$ for a predefined standard deviation $\sigma_s$. Consequently, $d_{shiftX}$ and $d_{shiftY}$ (in px) are added to the original positions of the box, shifting its position. For area perturbations, we increase/decrease its height and width by drawing $ d_{areaX} $ and $d_{areaY}$ independently for each annotation from a normal distribution: $ d_{areaX}, d_{areaY}  \sim \mathcal{N}(1, \sigma_a)$ for a predefined standard deviation $\sigma_a$. We use $d_{areaX}$ and $d_{areaY}$ to scale the original width and height of the annotation boxes while keeping its centroid fixed. After applying the perturbations, we ensure that the rectangles are fully in the frame by shifting them accordingly. By systematically varying the standard deviations $\sigma_s, \sigma_a$, we can assess the robustness of approaches.
To ensure fairness, all approaches share the same set of perturbations.

\begin{figure}
    \centering
    \includegraphics[width=1\linewidth]{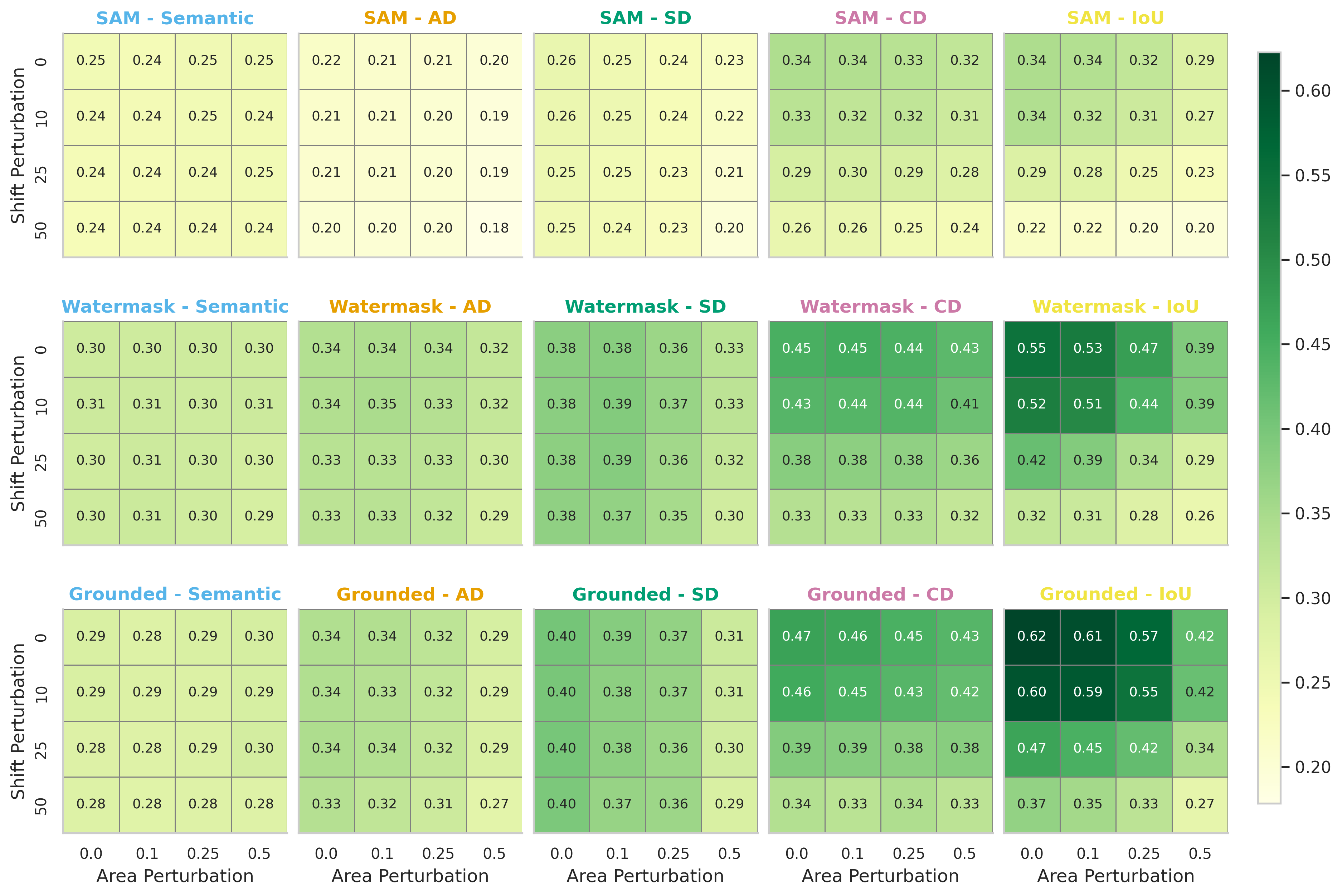}
    \caption{Heatmap showing R@100 for (model, rectangle-distance) pairs for skippable annotations.}
    \label{fig:perturbations_100}
\end{figure}

We show the results in \autoref{fig:perturbations_100} for $\sigma_a \in [0.0, 0.1, 0.25, 0.5]$  and  $\sigma_s \in [0,10,25,50]$ measuring $R@100$.  While all approaches are prone to perturbations, the intensities are varying. Most of the (model, distance) combinations that achieve high retrieval accuracy without perturbations also have higher accuracy with perturbations. An exception \GIOU\ (especially for the Grounded and Watermask models) that strongly suffers from both area and shift perturbations - this leads to the effect that its performance drops below \GCD, which is mostly robust to area perturbations. For high numbers of shift perturbations, all models fail, even \GAD\ and \GSD\ - leading to the conclusion that bounding boxes are not even matched with the fitting target regions anymore. However, the rectangle distances remain still more effective than all baselines, although they experience heavy performance drops. Compared to Jäckl et al.~\cite{staticApproach},  who also tested the robustness for static partitionings, rectangle-based distances are more prone to outliers but remain superior due to their high baseline accuracy.

%% file: Chapter/5-Conclusion.tex
This work investigated image retrieval effectiveness for dynamic sub-region search in homogeneous domains. Therefore, we build a retrieval pipeline that has two cornerstones: 1) we use region detectors to extract sub-regions automatically, which can then be retrieved via region-constrained queries. 2) we use a combination of semantic distances (through CLIP) and geometric distances (through the specified constraints in the annotation and the bounding box of extracted regions) for ranking. This combination is beneficial, strongly improving whole-image and static sub-region baselines. A combination of CLIP with an IoU-based distance is especially effective, doubling recall rates. We find that although combined distance functions are prone to perturbations in the annotation box, they maintain a higher accuracy than all baselines. However, our experimental setup is limited as the ground-truth annotations consisted of perfect bounding boxes, as the target image was directly presented to the user - those are thus unachievable in practice. Consequently, all presented results are theoretical upper bounds, highlighting the potential of dynamic sub-region search. Therefore, a self-evident future work is to collect annotations in a realistic setup. Another limitation of our experiments is that we investigated only one homogeneous domain. Thus, our approach is only transferable if accurate region detectors exist.